BRIEF RESEARCH REPORT

# Dual-Functional Cerium Oxide Nanoparticles with Antioxidant and DNase I activities to Prevent and degrade Neutrophil Extracellular Traps

**Hachem DICH†,1, Ramy ABOU RJEILY†,1, Gabriela RATH1, Mathéo BERTHET2, Bénédicte DAYDE-CAZALS2, Jean-François BERRET3, Eduardo ANGLES CANO*,1**

[1]*Faculté de Santé, Université Paris Cité, INSERM, Optimisation thérapeutique en neuropharmacologie, U1144, Paris, France.*
[2]*Specific Polymers, ZAC Via Domitia, 150 Avenue des Cocardières, 34160 Castries, France.*
[3]*Université Paris Cité, CNRS, Matière et Systèmes Complexes, 75013 Paris, France.*

† H.D. and R.A.R. These authors contributed equally to this work and share first authorship.
*Eduardo.Angles-Cano@inserm.fr

ABSTRACT
Neutrophils play a central role in immunothrombosis through the formation of neutrophil extracellular traps (NETs), a process known as NETosis. Upon stimulation, neutrophils release decondensed chromatin structures enriched with proteolytic enzymes, which contribute to thrombus formation. NETosis is critically dependent on reactive oxygen species (ROS), making redox regulation a key point of intervention. The intrinsic redox cycling of cerium oxide nanoparticles (CNPs) imparts self-regenerating antioxidant properties suitable for modulating neutrophil-driven oxidative stress. To address both the prevention and clearance of NETs, we developed dual-functional CNPs conjugated with DNase I. These engineered nanoparticles were efficiently internalized by neutrophils, reduced intracellular ROS levels, and inhibited NETs formation. In addition, DNase I-functionalized CNPs degraded pre-formed NETs. This dual-action strategy offers a promising nanotherapeutic platform for mitigating NETs-associated thrombotic pathologies. Ongoing studies aim to enhance thrombus targeting and assess *in vivo* efficacy.

## INTRODUCTION

Neutrophils are the first line of defense against invading pathogens, initiating a sequential response that includes degranulation, phagocytosis, and the release of neutrophil extracellular traps (NETs), a process known as NETosis (1-3). NETosis is considered the ultimate effector mechanism of neutrophils (4) that depends on the production of reactive oxygen species (ROS) by the nicotinamide adenine dinucleotide phosphate (NADPH) oxidase pathway. ROS trigger activation of Protein-Arginine deiminase 4 (PAD4) and the release of elastase from azurophil granules, leading to citrullination and degradation of histones and finally to chromatin decondensation. During this process, decondensed nuclear DNA is extruded into the extracellular space, complexed with histones and granular proteases such as human neutrophil elastase (HNE), cathepsin G, and myeloperoxidase, forming a filamentous web that captures pathogens (5, 6). Beyond their antimicrobial role, NETs have now been implicated in various sterile inflammatory disorders, including autoimmune diseases and thrombosis (5,7-10). NETs significantly influence coagulation by promoting platelet adhesion and activating clotting factors, contributing to thrombus formation in the microcirculation, (*i.e.*, septic shock *(11, 12))* and in medium size arteries (i.e., myocardial infarction (13, 14), and ischemic stroke (15)). The interplay of NETs with the coagulation system is central to the concept of immunothrombosis, a complex process in which immune defense mechanisms, particularly those involving NETs, contribute to blood clot formation (6, 11), with studies showing that NETs are present in 20.8% of thrombus in myocardial infarction (16). Furthermore, recent analyses of thrombus composition in ischemic stroke show that besides fibrin and blood cells, NETs are now recognized as substantial components of ischemic stroke thrombi underscoring their relevance in thrombotic disease (15). There is strong, converging evidence suggesting that the anti-thrombolytic effect attributed to neutrophils depends on the release of NETs (17, 18).

Crucially, NETs impair thrombus dissolution by interfering with fibrinolysis. The fibrinolytic system primarily targets fibrin, while NETs, composed largely of DNA, remain resistant to conventional thrombolytic agents (17), thus highlighting the need for alternative strategies to degrade NETs. Deoxyribonuclease I (DNase I) and DNase1-Like 3 play a dual-protection system to degrade intravascular NETs (19-21). DNase1L3 degrade chromatin and apoptotic microparticles. (22, 23) DNase I is known to predominantly target double-stranded extracellular DNA and has been shown to enhance thrombus lysis, supporting its potential use as an adjuvant to thrombolytic therapy (24, 25).

Since NETosis is driven by ROS, it may be prevented by targeting ROS production generated via the NADPH oxidase pathway. In this regard, cerium oxide nanoparticles (CNPs) have garnered interest for their potent antioxidant properties, attributed to their ability to cycle between $Ce^{3+}$ and $Ce^{4+}$ oxidation states, effectively neutralizing ROS (26, 27). Studies have demonstrated that CNPs exhibit significant antioxidant effects both *in vitro* and *in vivo*, making them promising candidates for mitigating oxidative stress in inflammatory diseases, including cancer, autoimmune disorders, and neurodegenerative conditions (28-31). For use in both *in vitro* and *in vivo* applications, CNPs must be functionalized to prevent protein corona formation and subsequent aggregation (32). A recent study demonstrated that CNPs can be effectively coated with functional copolymers using a two-step coating process, in which nanoparticles and polymers are synthesized separately and then assembled through non-covalent interactions (33). This method provides excellent colloidal stability across various solvents, including buffers and cell culture media (34). Importantly, it was confirmed that the superoxide dismutase (SOD)-like activity of the coated particles remains comparable to that of uncoated CNPs, while their catalase (CAT)-like activity is reduced by approximately 30%. In the present study, we apply the functionalization strategy previously developed by us for metal oxide nanoparticle coating (33).

In this study we propose an original approach integrating both, prevention and degradation of NETs. We investigated the ROS-scavenging potential and antioxidant effects of CNPs on human neutrophils and their effect on NETosis. Additionally, we explored a dual therapeutic approach by functionalizing CNPs with DNase I to degrade pre-formed NETs. To this end, we monitored neutrophil and NETs morphology, quantified ROS production, and analyzed the DNA and protein composition of NETs. Our aim was to determine whether CNPs, alone or conjugated to DNase I, can be used to prevent NETosis and as an adjuvant therapy to improve thrombolysis and restore vascular flow.

**MATERIAL AND METHODS**

**Neutrophil isolation and generation of neutrophil extracellular traps**

Venous blood from healthy volunteers was collected on EDTA (23 mM) by the French Blood Donor Center. Neutrophils were isolated from the anticoagulated blood following a cell exclusion method (MACSxpress®, Miltenyi Biotec) (35). Non-target cells were removed by immunomagnetic depletion using magnetic beads coated with specific antibodies directed against all blood cells but neutrophils. Flow cytometry analysis confirmed the recovery of CD16+ (Fc receptor, FcgRIII) highly purified neutrophils (≥ 98%) that were identified by cell viability (Trypan blue) and nuclear morphology on May-Grünwald-Giemsa staining. Isolated neutrophils were resuspended in Hank's balanced salt solution (HBSS) and labelled with the fluorescent cell linker dye PKH26, (Sigma-Aldrich), which intercalates into the phospholipid bilayer of the neutrophil membrane due to its lipophilic nature. PKH26-labelled neutrophils were used immediately after isolation at a concentration allowing distribution of 100,000 cells per well on 96-well flat-bottomed plates. Plated neutrophils were treated with 50 nM phorbol-12-myristate-13-acetate (PMA, Sigma-Aldrich) in a humidified 5% $CO_2$ atmosphere at 37 °C for 4 hours. NETs were identified by (1) DNA staining with a 10 µg/ml solution of Hoechst 33342 (Invitrogen) and (2) measuring the activity of DNA-bound elastase as indicated below. Stained non-stimulated neutrophils and PMA-induced NETs were detected in optic fields of three wells for each condition using the x10 objective of a Zeiss AxioObserver D1 fluorescence microscope equipped with a CCD Imaging camera using the Histolab software from Microvision Instruments (Evry, France). Neutrophil membrane labeled with PKH26 was visualized with the Rhodamine filter (excitation ⊠551nm, emission ⊠567nm). DNA was visualized using a 4', 6-diamidino-2-phenylindole (DAPI) filter (excitation $\lambda$350nm, emission $\lambda$461nm) (Figure 1).

**Characterization of NETs**

NETs are typically composed of DNA fibres, histones and granular proteases, primarily elastase.[5] DNA, the core component of NETs was quantified using SYBR Green (Invitrogen, $\lambda_{ex}$=497nm, $\lambda_{em}$=520nm). For this purpose, we constructed a reference range (0.05 to 10 to µg/ml) of fibrillar DNA (Sigma-Aldrich® D1501) diluted in HEPES buffer (HEPES 10mM, Tween 20 0.05%, NaCl 140mM, pH 7.4). An equal volume of SYBR Green (1:5000) in HEPES buffer was added and after 15min incubation in the dark at room temperature, the fluorescence was measured using a FLX800 microplate fluorescence reader (Bio-Tek Instruments). Elastase, another major component of NETs, was quantified using a reference range of 6.25 to 100 nM human neutrophil elastase (HNE, Enzo®) diluted in HEPES buffer. The chromogenic substrate *N*-Methoxysuccinyl-Ala-Ala-Pro-Val-*p*-nitroanilide (Merck®) (100 µl per well, 1.5 mM final concentration in HBSS) was incubated with NETs after supernatant removal. DNA-bound HNE activity was detected at 37 °C by measuring the release of *p*-nitroaniline ($\lambda$ $A_{405nm}$/$\lambda$ $A_{490nm}$) in a microplate reader FlexA-200 (Allsheng-Instruments) by following the reaction kinetics for 1h at 37°C. The elastase inhibitors MeO-Suc-AAPV-CMK, Merck®, (100 µM final) or $\alpha_1$-proteinase inhibitor (Sigma-Aldrich) (10 µM final) and 4-(2-aminoethyl)benzene-sulfonyl fluoride (AEBSF, Sigma-Aldrich) (1 mM) were used to authenticate preservation of DNA-bound elastase activity.

**Synthesis and PEG-based surface functionalization of CNPs**

CNPs were synthesized by thermohydrolysis of cerium(IV) nitrate under acidic conditions, yielding nanocrystals with a median diameter of 7.8 nm and a crystallite size of 2.6 nm (26, 36). The uncoated particles exhibit a positive surface charge ($\zeta$ = +40 mV) and a $Ce^{3+}$ content of 20%, ensuring colloidal stability in acidic media (37). To improve dispersion stability and biofunctionality, the CNPs were coated with phosphonic acid-terminated polyethylene glycol (PEG) polymers synthesized via free radical polymerization (33). Two types were used: a homopolymer (PEG2ka-Ph) with a primary amine, and a random copolymer (MPEG2k-co-MPh) containing multiple phosphonic acids for stronger anchoring. Coated particles (CeO2@PEG2ka-Ph and CeO2@MPEG2k-co-MPh) were stable in physiological media, with coating layers of 5–10 nm (34). For tunable surface charge, hybrid coatings combining both polymers at different ratios were also prepared. All CNPs coated with both polymers exhibited good colloidal stability. Finally, to enable continuous modulation of the surface cationic charge on PEG-coated particles, favoring interaction with the plasma membrane of neutrophils, we used a mixture of 30% $MPEG_{2k}$-co-MPh and 70% $PEG_{2k}$a-Ph to coat the nanoparticles, yielding hybrid structures denoted $CeO_2@(MPEG_{2k}\text{-co-MPh})_{0.3}(PEG_{2k}a\text{-Ph})_{0.7}$. Details for coating protocol and coated CNPs final preparation steps are provided in Supplementary Material (Supplemental Method 1 Coating protocol; Supplemental Method 2 Polymer coating). The general physicochemical properties of the three types of CNPs used in this study and transmission electron microscopy of CNPs bearing different coatings are provided in Supplementary Table II and Table III, respectively.

**Grafting of DNase I onto CNPs ($CeO_2@(MPEG_{2k}\text{-co-MPh})_{0.3}(PEG_{2k}a\text{-Ph})_{0.7}$)**

DNase I (Pulmozyme®, Roche; 1 mg/1000 U/mL) was grafted onto $CeO_2@(MPEG_{2k}\text{-co-MPh})_{0.3}(PEG_{2k}a\text{-Ph})_{0.7}$ nanoparticles using polyglutaraldehyde as a linker. A 2.5% glutaraldehyde solution (prepared from a 25% stock, TAAB Labs) was diluted in sodium bicarbonate buffer (pH 9.5) and incubated at 37°C for 15 minutes. Polymer formation was monitored by measuring the monomer-to-polymer absorbance ratio at $\lambda A_{285nm}/\lambda A_{233nm}$, with an acceptable threshold of $\leq 0.1$ (Nanodrop One, Thermo Scientific). The polyglutaraldehyde solution was then incubated with coated CNPs (10 mg/mL) for 15 min at room temperature to facilitate polymer immobilization. Unbound polyglutaraldehyde was removed using Amicon® Ultra 30K centrifugal filters. The resulting polyglutaraldehyde-activated CNPs (10 mg/mL) were then incubated with DNase I (1 mg/mL) for 1 hour at 22°C. Excess unbound DNase I was removed by ultrafiltration, followed by a final wash. The grafted CNPs were resuspended in 10 mM HEPES buffer, pH 7.4, containing 0.14 M NaCl, 5 mM $CaCl_2$, and 5 mM $MgCl_2$. Polyglutaraldehyde acted as a spacer, preserving the conformation and enzymatic activity of DNase I by preventing direct interaction with the CNP surface and coating. DNase I grafting efficiency was assessed using a Bradford protein assay (Merck). The amount of DNase I enzymatic activity grafted onto $CeO_2@(MPEG_{2k}$-co-$MPh)_{0.3}(PEG_{2k}a$-$Ph)_{0.7}$ was quantified using a DNase I reference curve (0–100 µg/mL) in a DNA degradation assay, where fibrillar DNA served as the substrate to mimic NETs. CNPs grafted with DNase I were tested at three concentrations (25, 50, and 100 U/mL), and their activity on fibrillar DNA and native NETs was directly compared to that of equivalent concentrations of free DNase I.

**Adsorption of dyes to $CeO_2@(MPEG_{2k}$-co-$MPh)_{0.3}(PEG_{2k}a$-$Ph)_{0.7}$ CNPs**

Interestingly, besides their catalytic effectiveness, ceria has excellent dye adsorption capacity (38, 39). We tested the hypothesis that dye adsorption onto CNP could be useful to detect phagocytosis by neutrophils. We used two fluorescent dyes: 3,3'-Dihexyloxacarbocyanine Iodide (DiOC6, $\lambda_{ex}$=484 nm, $\lambda_{em}$=501 nm) and Rhodamine 6G (RH6G, $\lambda_{ex}$=528 nm, $\lambda_{em}$=548 nm) in order to follow the CNPs phagocytosis by neutrophils. To incorporate the dyes, $CeO_2@(MPEG_{2k}$-co-$MPh)_{0.3}(PEG_{2k}a$-$Ph)_{0.7}$ was incubated with 1/20th and 1/100th dilutions of DiOC6 and RH6G, respectively, for 30min at room temperature, protected from light. To wash off the dye excess, the mixture was centrifuged using 30kDa ultrafiltration tubes at 13,000rpm, for 5 min at room temperature, 4 times, each time adding 100µl of HBSS. The retentate from the last wash was recovered, and topped up to the initial volume to preserve the initial cerium oxide concentration. Alternatively, we grafted $CeO_2@(MPEG_{2k}$-co-$MPh)_{0.3}(PEG_{2k}a$-$Ph)_{0.7}$ with a commonly used fluorophore, Cyanin-5 (Cy-5), (see Supplementary Figure 4) to verify efficacy of DiOC6-CNP internalization (Figure 2A). PEGylation forms a barrier around the core, preventing non-polar molecules from crossing the shell. DiOC6 is an amphiphilic cationic dye (lipophilic alkyl tail and positively charged carbocyanine)(40) that interact with the hydrophilic PEG layer and penetrate the brush-like architecture allowing efficient adsorption onto the cerium surface.

**Treatment of neutrophils with DNase I-CNPs**

Neutrophils were seeded at varying amount ($5x10^4$ to $15x10^4$) of cells per well in a 96-well microtitration plate. Seeded neutrophils were tested under three conditions: 1) HBSS buffer, 2)

N-acetylcysteine (NAC) used as a positive antioxidant control at 10μM or 3) DiOC6-labelled-CNPs. For all conditions, incubation was allowed for 30 min before the addition of 50 nM PMA or HBSS buffer, after which a second incubation was performed for 4 h in a 37 °C humidified atmosphere of 5% $CO_2$. PEGylated CNPs were added at 4 different concentrations (100, 250, 500 and 1000 μg/ml). To observe and characterize the neutrophil phenotype after these treatments, a Zeiss Axio Observed fluorescence microscope was used at x10 magnification. Neutrophil membrane labeled with PKH26 was visualized with the Rhodamine filter. DNA was visualized using Hoechst33342 and a DAPI filter. DiOC6 labeled- and RH6G labeled-CNPs were detected using the Endow GFP and the Rhodamine filters respectively. The colocalization of fluorescent markers was analyzed by merging the images and quantification by using ImageJ and Prism softwares.

**Measurement of ROS**

ROS produced by stimulated neutrophils were quantified using the fluorescent dye dichlorodihydrofluorescein diacetate ($H_2$DCFDA, ThermoFischer-Invitrogen) ($\lambda_{ex}$=492nm, $\lambda_{em}$=517nm) (32). The dye, dissolved extemporaneously with 50 μl of dimethylsulfoxide (DMSO), was added to a suspension of neutrophils treated under different conditions at a 10 μM final concentration. Neutrophils were incubated in a 96-well microplate at a concentration of $1 \times 10^5$ cells/well. The CNPs were added at a final concentration of 1000 μg/ml, whereas the concentration of the grafted particles was expressed in U/ml on the basis of the amount of grafted DNase I. Conditions to stimulate the production of ROS were as described above. Fluorometer readings were taken every 5 min for 2 h at 37°C using a VARIOSKAN-LUX (Thermo Scientific).

**Statistical analysis**

Quantified data are presented as mean ± SD and were compared using an ordinary one-way ANOVA. GraphPad Prism version 10.2.3 for Windows was used for the analyses. Differences were considered significant with a *p*-value<0.05.

## RESULTS

**Stimulation of neutrophils by PMA causes the transmutation of chromatin into NETs**

We employed the well-established method of PMA stimulation to induce NETs, as PMA, similar to bacteria and IL-8, activates a common signaling cascade involving protein Kinase C and the generation of ROS (41). As expected, PMA-stimulated neutrophils produced robust NETs, characterized by extensive chromatin decondensation leading to the formation of extracellular DNA fibers (Figure 1B). In contrast, neutrophils pre-treated with NAC, the antioxidant positive control, and subsequently stimulated with PMA retained their characteristic polylobed nuclear morphology, indicating a resting state (Figure 1C). Interestingly, a subset of neutrophils incubated in buffer alone displayed decondensed nuclei that occupied the entire cytoplasm, hereafter referred to as "cloudy" cells, suggesting low-grade activation potentially caused by handling or experimental manipulation (Figure 1A).

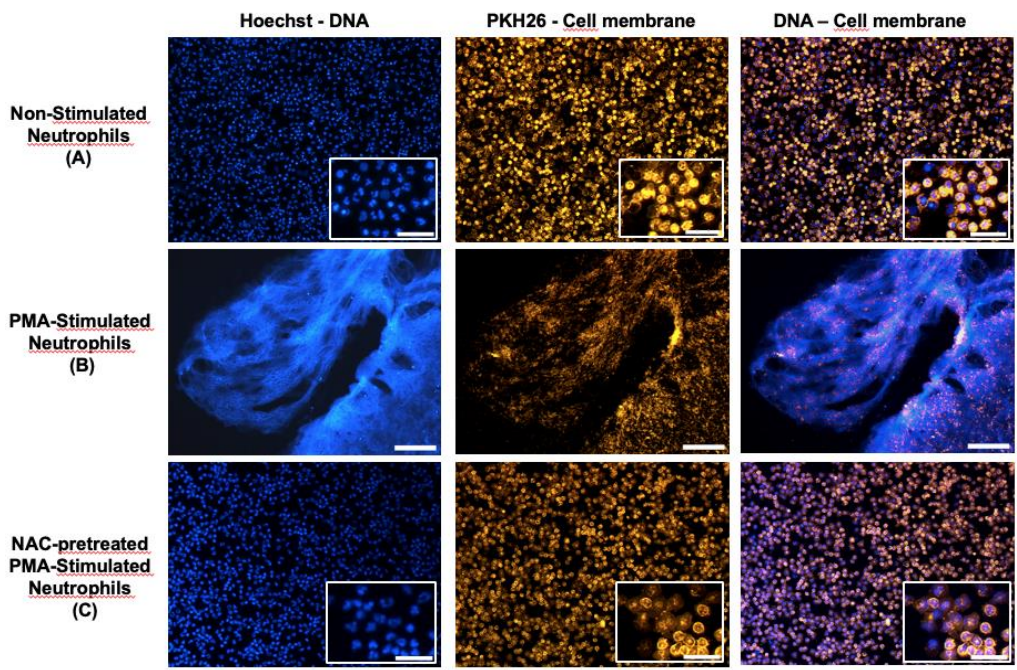

**Figure 1.  Phenotypic characterization of neutrophils**
*Neutrophils isolated from human blood were seeded on a 96-well plate at 1 x $10^5$ cells/well and treated under different conditions. n=3 (**A**) Non-stimulated neutrophils, incubated with buffer (HBSS) alone. (**B**) Neutrophils stimulated with PMA (50µM), scale bar 150 µm. (**C**) Neutrophils treated with the positive antioxidant control N-acetyl-l-cysteine (NAC, 10µM) before PMA (50nM) stimulation. Representative images of fluorescence microscopy: Left column: Hoechst 33342 for DNA staining. Middle column: PKH26 staining for membrane labelling. Right column: merged left and middle raw images.  Insets (**A** and **C**): High magnification images (500%), scale bar 50 µm. Images were obtained from optic fields of three wells for each condition using the x10 objective of a Zeiss AxioObserver D1 fluorescence microscope equipped with a CCD Imaging camera.*

**Characterization of NETs by DNA and Elastase content**

Results obtained using neutrophils isolated from different healthy donors used in these experiments are shown in Supplementary Figure 3. A dose-dependent decrease in detectable DNA was observed with increasing DNase I-grafted CNPs concentrations (360 µg/ml•100 U/mL and 180 µg/ml•50 U/mL) in both wells (A) and supernatants (B) (Supplementary Figure 3A). The highest DNA levels were detected in samples treated with the lowest DNase I-grafted concentration (18 µg/ml•5 U/mL), indicating incomplete degradation at the well bottom with release of longer cell-free DNA (cfDNA) fragments into the supernatant (Supplementary Figure 3B). The partial degradation of NETs, generates smaller DNA fragments that are released into the supernatant. These smaller fragments remain soluble after centrifugation and are efficiently detected by SYBR Green, thus explaining the high cfDNA detected in Supplementary Figure 3B at 5 U/mL DNase I. In contrast, at higher DNase I concentrations (50–100 U/mL), a more complete digestion produces fragments that are either too small to be stained effectively by SYBR Green or are degraded beyond detection.

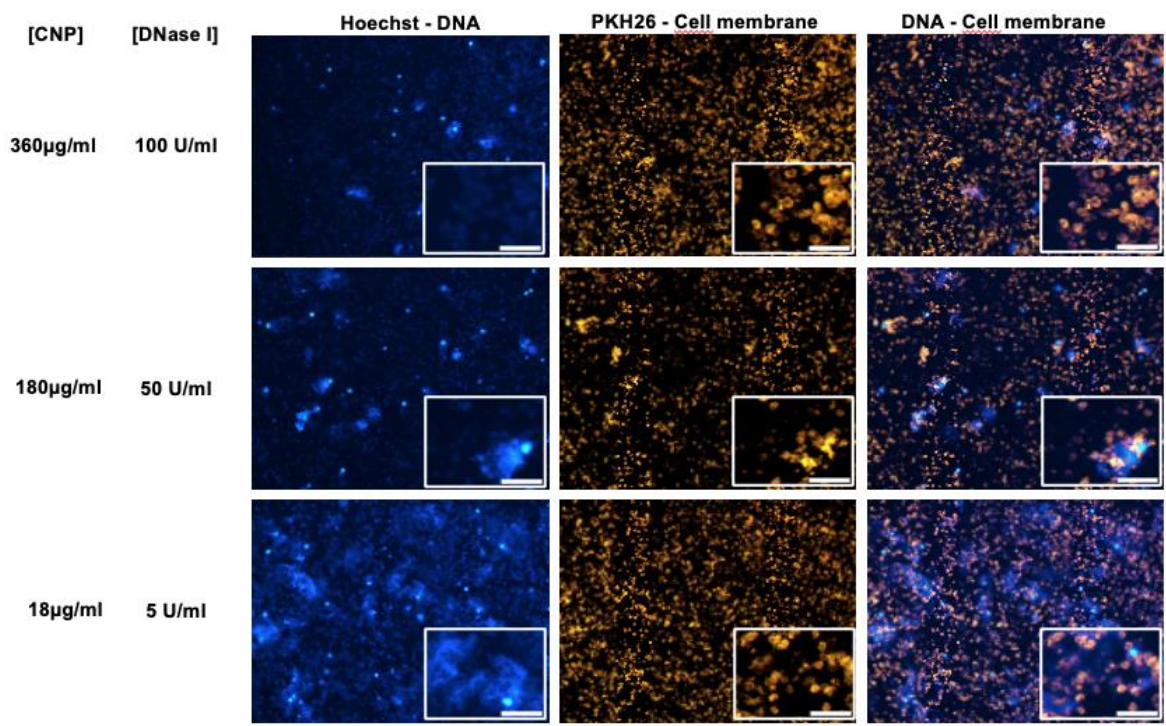

**Figure 3. Impact on neutrophil DNA of CeO2@(MPEG2k-co-MPh)$_{0.3}$(PEG2ka-Ph)$_{0.7}$ grafted with DNase I**

*Different concentrations of DNase I grafted on CeO2@(MPEG2k-co-MPh)$_{0.3}$(PEG2ka-Ph)$_{0.7}$ were incubated with neutrophils 30 min prior to stimulation with 50 nM PMA (see Figure 1 for con-ditions) n=6. Representative fluorescent microscopy images of neutrophils stained and cap-tured as indicated in Figure 1. Insets: High magnification (500 %) images of polylobate neutro-phils (scale bar 50 µm) at three different concentrations of CeO2@(MPEG2k-co-MPh)$_{0.3}$(PEG2ka-*

*$Ph)_{0.7}$. The indicated concentrations of nanoparticles 360, 180, and 18µg/ml correspond, respectively, to 100, 50 and 5 U/ml grafted DNase I. Cytoplasmic decondensed DNA of cloudy cells decrease as a function of the concentration of grafted DNase I.*

Panels C and D show HNE levels measured in the wells bottom and in the supernatants, respectively. As shown in Figure 3 (Hoechst-DNA column), no NETs were observed at high concentrations (360 µg/ml•100 U/mL and 180 µg/ml•50 U/mL) of CNP/DNase I (*Fig. 3, upper and middle panels*). As a consequence, the small amount (*ca* 50 nM) of HNE remaining at the bottom of the corresponding wells was most probably bound at the neutrophil membrane (Supplementary Figure 3C). However, in the corresponding cell-free supernatants (Supplementary Figure 3D), the DNase I-dependent increase in elastase concentrations reflects the release of DNA-bound elastase from degraded DNA fragments. However, at 18 µg/mL•5 U/ml, CNP•DNase I, elastase levels were lower because 5 Units of DNase I were insufficient to degrade NETs formed at 18 µg\ml CNPs.

**$CeO_2@PEG_{2k}$a-Ph nanoparticles effectively prevent NETosis**

We tested CNPs coated with various PEG polymers to identify formulations that retain their antioxidant activity in the presence of neutrophils (Supplementary Figure 5). Notably, no NET formation was observed when PMA-stimulated neutrophils were incubated with cationic $CeO_2@PEG_{2k}$a-Ph (Supplementary Figure 5A). Most neutrophils remained in the "cloudy cell" state, resembling the phenotype of non-stimulated controls, suggesting that the antioxidant capacity of $CeO_2@PEG_{2k}$a-Ph effectively prevented NETosis. In contrast, neutrophils treated with the uncharged $CeO_2@MPEG_{2k}$-*co*-MPh and positively charged $CeO2@MPEG2k$-co-MPEG2ka-co-MPh nanoparticles underwent typical NETosis (Supplementary Figure 5B and Figure 5C). To exert their enzymatic antioxidant mimetic activity, CNPs were phagocytosed by neutrophils, as shown by observations made using DiOC6-labeled CNPs (enlarged areas of Figure 2B). The quantification (mean±SD) of CNPs internalized by neutrophils is shown in Figure 2C. In all subsequent experiments, we assessed the efficacy of these dual-polymer coated CNPs, combining $MPEG_{2k}$-*co*-MPh and $PEG_{2k}$a-Ph at a 30:70 ratio. These CNPs were additionally functionalized with DNase I, as described in the Methods section. To evaluate dose-responsiveness, we prepared CNPs at final concentrations corresponding to 360 µg/mL, 180 µg/mL, and 18 µg/mL (Figure 3) respectively conjugated to DNase I at the concentrations 100U, 50U and 5U.

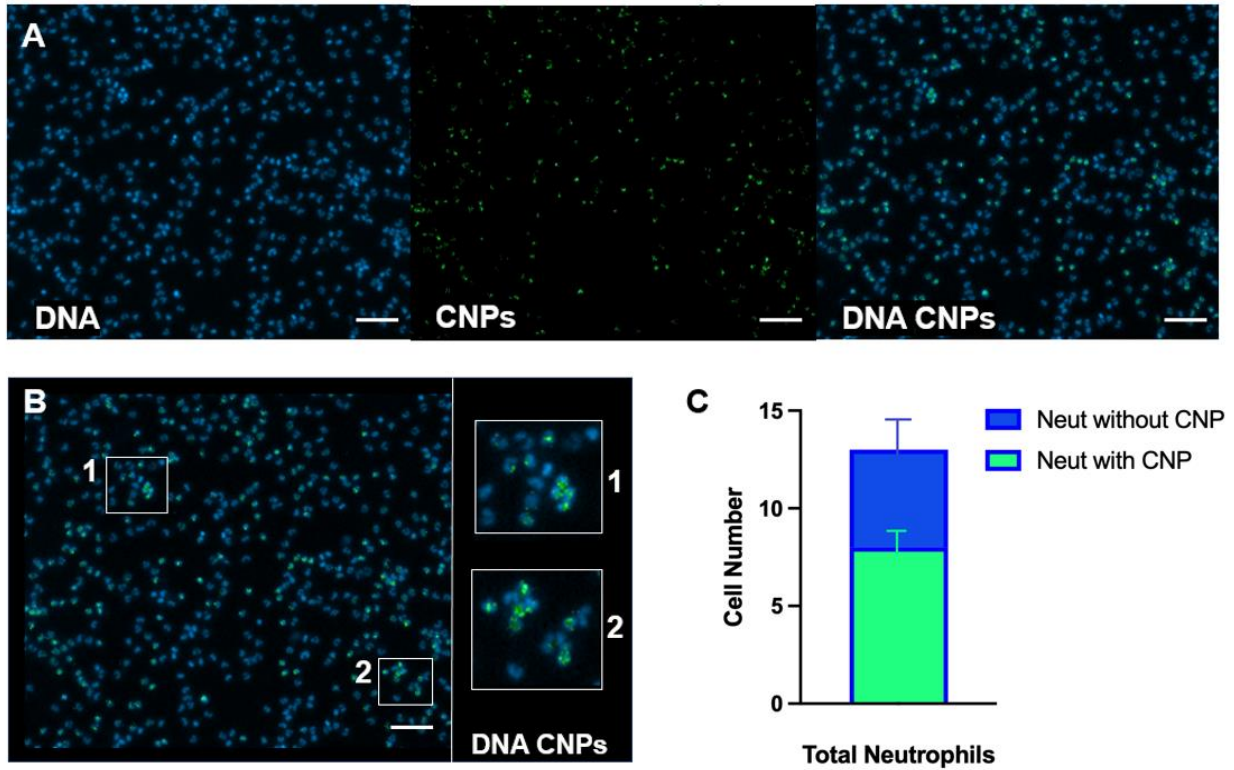

**Figure 2 . Phagocytosis of fluorescently labelled CNPs by neutrophils.**
*The fluorescent dye DiOC6 was incorporated on CeO2@(MPEG2k-co-MPh)$_{0.3}$(PEG2ka-Ph)$_{0.7}$ by chemisorption (pseudo-second-order kinetic reaction) as indicated in Methods. Neutrophils were seeded, stained and captured by fluorescent microscopy as indicated in Figure 1. Prior to stimulation with 50 nM PMA the neutrophils were treated with N-acetyl-l-cysteine (NAC, 10μM, positive control) and DiOC6-labelled CeO2@(MPEG2k-co-MPh)$_{0.3}$(PEG2ka-Ph)$_{0.7}$ (500 μg/ml). n=2 (A) Left panel: representative fluorescence images of neutrophils stained with Hoechst 33342 for DNA. Middle panel: DiOC6 adsorbed to CeO2@(MPEG2k-co-MPh)$_{0.3}$(PEG2ka-Ph)$_{0.7}$. Right panel: internalized CNPs by neutrophils (merged images). Scale bar: 100 μM. (B) Zoom (200%) of merged pictures shown in Figure 2A right panel. Scale bar 100 μM. (1) and (2) are zoomed fields clearly showing CeO2@(MPEG2k-co-MPh)$_{0.3}$(PEG2ka-Ph)$_{0.7}$ internalized by neutrophils (scale bar 25 μM). (C) Quantification from three fields of (B) showing the mean total neutrophil count (blue) and the mean number of neutrophils with phagocytosed CNPs (green). Error bars: SD.*

## DNase I-grafted CNPs degrade NETs in a dose-dependent manner

DNA degradation by DNase I grafted onto CNPS was comparable to that obtained with equivalent concentrations of free DNase I, the latter showing slightly higher, but not significant, activity (2.5 μg/mL fibrillar DNA treated with 100U/mL DNase I: 0.15 μg/ml vs 0.27 μg/mL residual DNA for soluble vs CNP-grafted form, respectively). A clear dose-response effect was observed with DNase I-functionalized CeO2@(MPEG2k-co-MPh)$_{0.3}$(PEG2ka-Ph)$_{0.7}$ on NET prevention and degradation (Figure 3). At concentrations of 360 μg/ml•100 U/ml and 180 μg/ml•50 U/mL, of CNP/DNase I nanoparticles effectively prevented NETosis and degraded cytoplasmic decondensed DNA of cloudy cells *i.e.* neutrophils at the chromatin decondensed state, as a function of the concentration of grafted DNase I in a dose-dependent manner (Figure 3 upper and middle panels). Whereas the functionalized CNPs bearing 18 µg/ml•5 U/mL of DNase I had no effect on NET prevention thus allowing their formation and partially degraded DNA of cloudy cells (Figure 3 lower panel). Neutrophils in the "cloudy" state were particularly susceptible to 100 and 50 U/ml of DNase I-mediated digestion (Figure 3 Hoechst-DNA column, upper and middle panel) as compared to 5U/ml DNase I (Figure 3 Hoechst-DNA column, lower panel). Neutrophils exposed only to the antioxidant effect of PEGylated CNPs remained mostly in the cloudy phase (Supplementary Figure 5A).

## Quantification of ROS produced by neutrophils

Given the pivotal role of ROS in NET formation, we quantified their production kinetics in neutrophils using fluorimetry (Figure 4). PMA-stimulated neutrophils exhibited the steepest kinetic curve (Figure 4A), reflecting the highest rate of ROS generation. The antioxidant effect of various treatments was captured by measuring $Vi_{max}$ (AU/min), allowing clear discrimination from the basal ROS production observed under non-stimulated conditions (HBSS) (Figure 4B). Interestingly, even under non-stimulated conditions, a slight increase in ROS was detected, likely reflecting low-grade activation due to experimental handling, thus explaining the finding of "cloudy cells" (Figure 1A, see decondensed DNA occupying the intracellular space). Fluorescence curves obtained from neutrophils treated with NAC, the positive antioxidant control, remaining near baseline a finding in agreement with the preservation of polylobate nuclei (see Figure 1C).

The effect of DNase I-functionalized $CeO_2@(MPEG_{2k}\text{-}co\text{-}MPh)_{0.3}(PEG_{2k}a\text{-}Ph)_{0.7}$ on ROS production was dose-dependent. Importantly, since the amount of grafted DNase I was directly proportional to the quantity of nanoparticles administered, higher DNase I concentrations reflect higher cerium oxide doses. Accordingly, the observed inverse relationship between DNase I concentration and the slope of the ROS generation curve reflects the increasing antioxidant effect of functionalized CNPs, rather than a direct effect of DNase I on ROS suppression.

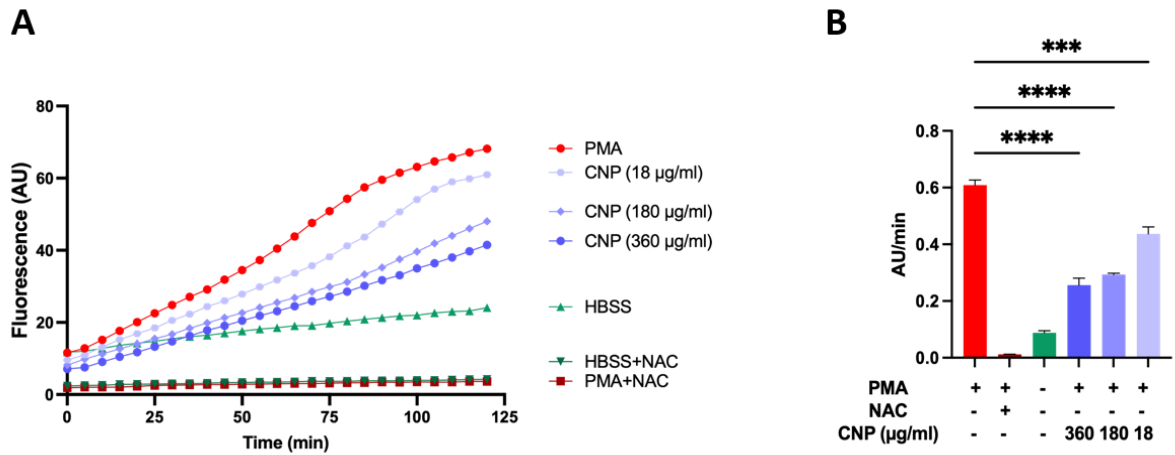

**Figure 4. Kinetics of ROS produced by neutrophils treated with grafted CeO2@(MPEG2k-co-MPh)$_{0.3}$(PEG2ka-Ph)$_{0.7}$.**

*ROS production by neutrophils was detected with the fluorescent dye dichlorodihydrofluoresce-in diacetate (H2DCFDA). The graph shows representative curves of ROS total production by neutrophils obtained under different CNP CeO2@(MPEG2k-co-MPh)$_{0.3}$(PEG2ka-Ph)$_{0.7}$ treatment conditions and detected using the fluorescent dye H2DCFDA. Red circles: neutrophils stimulated with PMA (50nM). Brown squares: neutrophils treated with the positive antioxidant control NAC (10µM) prior to PMA (50nM) stimulation. Green triangles: neutrophils incubated in HBSS buffer. Purple circles: neutrophils treated with nanoparticles grafted with DNase I (100U/ml, 50U/ml, and 5U/ml darkest to lightest shade of purple) prior to stimulation with 50nM PMA. (B) Bars represent max Vi calculated as fluorescence AU/min for each of the kinetics curve represented in (A). PMA: phorbol 12-myristate 13-acetate. NAC: N-acetyl-l-cysteine. HBSS: Hank's balanced salt solution. CNP: CeO2 nanoparticle. Error bars represent SD, n=4. Statistical significance is indicated by asterisk ****P<0.0001, ***P<0.001.*

## DISCUSSION

In this study, we demonstrate that the formation of NETs can be effectively impaired by the antioxidant superoxide dismutase- and catalase-mimetic activity of CNPs coated with PEG-terpolymers and functionalized for cellular uptake. These CNPs act as potent scavengers of ROS, thereby targeting a central driver of NET formation. ROS production is a key step in the NETosis pathway and is primarily mediated by NADPH oxidase, an oxidoreductase that generates ROS (4). One of the most widely used NET-inducing agents in experimental systems is PMA, a potent activator of NADPH oxidase. PMA stimulation mimics physiological triggers such as bacterial products and IL-8 by activating protein kinase C, which in turn initiates the Raf–MEK–ERK signaling cascade, culminating in NADPH oxidase activation and robust ROS generation.

Given the overlap in signaling pathways between PMA, bacterial stimuli, and inflammatory cytokines, we employed PMA as a well-standardized model to assess NET formation and to evaluate the capacity of CNPs to suppress this response via ROS inhibition (41). Building on this mechanistic understanding, we propose targeting ROS as a strategy to inhibit their production in activated neutrophils. CNPs are well recognized for their potent antioxidant properties, including catalase- and superoxide dismutase-mimetic activities (26). Leveraging this redox behavior, we employed CNPs to suppress ROS production in PMA-stimulated neutrophils (28, 29). The antioxidant activity of CNPs is primarily attributed to cerium's ability to reversibly cycle between $Ce^{3+}$ and $Ce^{4+}$ oxidation states (26). This redox cycling enables a self-regenerating mechanism that dynamically responds to oxidative stress, allowing sustained ROS scavenging both *in vitro* and *in vivo* (42).

In addition to their chemical robustness, CNPs display biomimetic properties that align with key antioxidant enzymes involved in cellular ROS regulation, including those in the NADPH oxidase pathway. These characteristics have prompted increasing interest in CNPs as therapeutic agents for a range of oxidative stress-related diseases, including cancer, autoimmune disorders, and neurodegenerative conditions such as Alzheimer's disease (29-31). In addition to evaluating their antioxidant activity, CNPs were functionalized with DNase I using a glutaraldehyde-mediated coupling strategy. This method forms imine bonds between the amine groups of DNase I and the aldehyde groups of glutaraldehyde, preserving the protein's conformational integrity and enzymatic activity. Our results showed that DNA degradation by the grafted DNase I was comparable to that of free DNase I, with the latter exhibiting slightly higher, although not significant, activity.

It is well-established that DNase I degrades NETs-DNA and is therefore therapeutically promising in conditions where NETs contribute to their pathophysiology (sepsis, myocardial infarction, and ischemic stroke.(12, 14, 15) Recent studies have shown that DNase I can accelerate the lysis of human coronary thrombi containing NETs, providing proof of concept that DNase I can be used as an adjuvant to thrombolytic therapy (13). However, the clinical utility of DNase I is limited by its short half-life and susceptibility to plasma proteases. To overcome these challenges, strategies such as surface grafting or encapsulation of DNase I into polymeric nanoparticles have been employed to protect the enzyme and extend its stability in circulation.(43, 44) In the present study we propose that beyond these protective strategies, the use of ter-poymer antioxidant CNPs offers a unique advantage by scavenging ROS and thereby preventing NETosis.

Our approach uniquely combines DNase I delivery alongside the intrinsic antioxidant properties of CNPs, which exhibit catalase- and superoxide dismutase-mimetic activities. This dual functionality - enzymatic degradation of NETs via DNase I and ROS scavenging via CNPs - distinguishes our platform from previously reported delivery systems. While alternative formulations, such as those recently described have demonstrated efficient DNase I delivery using lipid- or polymer-based nanoparticles. (43, 45, 46) These studies primarily target nuclease delivery without addressing oxidative stress, a key contributor to inflammation and NET-associated pathology. In contrast, our CNP-based system offers a multimodal therapeutic strategy that may enhance efficacy in diseases where both NET accumulation and oxidative stress drive pathology.

Following functionalization, $CeO_2@(MPEG_{2k}$-*co*-$MPh)_{0.3}(PEG_{2k}$a-$Ph)_{0.7}$ coupled with DNase I retained both their nuclease activity, effectively degrading fibrillar DNA and NETs, and their intrinsic antioxidant enzyme-mimetic properties. Importantly, to exert their antioxidant effect, the functionalized CNPs must be internalized by neutrophils. To monitor this uptake, we leveraged the high adsorption capacity of CNPs, a property widely exploited in environmental applications such as water decontamination. We employed fluorophores such as DiOC6 that, via its hydrophilic groups pass through the PEGylated shell, and adsorb onto CNP interfaces following a pseudo-second-order kinetics consistent with Langmuir-type adsorption models, enabling precise tracking of nanoparticle internalization (39).

In summary, we demonstrate that CNPs effectively scavenge ROS, thereby preventing NET formation and arresting neutrophils at the chromatin decondensation stage ("cloudy cells"). For already formed NETs, DNase I-functionalized CNPs provide an efficient mechanism for extracellular DNA degradation. These bifunctional nanoparticles thus offer a dual-action approach: preventing NETosis upstream through redox modulation, and dismantling established NET structures via nuclease activity. Given their combined antioxidant and DNase I-mediated properties, we propose these engineered CNPs as promising adjuvants to thrombolytic therapy. By targeting both the initiation and persistence of NETs within thrombi, $CeO_2@(MPEG_{2k}$-*co*-$MPh)_{0.3}(PEG_{2k}$a-$Ph)_{0.7}$-DNase I particles may synergize with recombinant tissue plasminogen activator (r-tPA), facilitating more effective degradation of NET-associated thrombi. This strategy could address a key limitation of current thrombolytics, which are often hindered by the structural stability and procoagulant nature of NETs. Future studies should explore the *in vivo* biodistribution, pharmacodynamics, and immunomodulatory profile of these multifunctional nanoparticles to assess their translational potential in thromboinflammatory conditions. While our findings highlight the therapeutic promise of bifunctional CNPs, several considerations remain before clinical translation. The long-term biocompatibility and clearance of cerium nanoparticles require thorough evaluation, particularly in the context of repeated or systemic administration. Additionally, while *in vitro* results are encouraging, *in vivo* models of thromboinflammation will be critical to assess efficacy, biodistribution, and potential off-target effects. Beyond thrombosis, the dual ability to modulate oxidative stress and dismantle extracellular DNA structures, positions these nanoparticles as potential candidates for diseases characterized by NET-mediated pathology, including myocardial infarction, ischemic stroke, and COVID-19-associated coagulopathy. These avenues merit further investigation to fully harness the clinical potential of CNP-based nanotherapeutics.

## Nonstandard abbreviations

AEBSF, 4-(2-Aminoethyl)benzenesulfonyl fluoride
HBSS, Hank's balanced salt solution
HNE, human neutrophil elastase
$H_2$DCFDA : 2',7'-dichlorodihydrofluorescein diacetate
NETs, neutrophil extracellular traps
MeO-Suc-AAPV-CMK, N-(Methoxysuccinyl)-L-alanyl-L-alanyl-L--prolyl-L-valine chloromethylketone
MeO-Suc-AAPV-*p*NA, N-(Methoxysuccinyl)-L-alanyl-L-alanyl-L--prolyl-L-valine *p*-nitroanilide
PMA, phorbol 12-myristate 13-acetate
PMSF, phenylmethylsulphonyl fluoride
DiOC6, 3,3'-Dihexyloxacarbocyanine iodide

## Acknowledgments

EA gratefully acknowledge the late Professor Michel Plotkin for his valuable insights during the early conceptualization of this study.  We would like to acknowledge Dr. Divina El Hamaoui and Bachelor student Josselin Bertrand for their assistance with data acquisition.  This work was funded in part by core support funds from INSERM and by the French National Research Agency grants ANR-20-CE18-0022-01 (project STRIC_ON) and ANR-23-CE17-0023-03 (project DELIASE).

# SUPPLEMENTARY MATERIAL

## Dual-Functional Cerium Oxide Nanoparticles with Antioxidant and DNase-I activities to prevent and degrade Neutrophil Extracellular Traps

Hachem DICH[§,1], Ramy ABOU RJEILY[§,1], Gabriela RATH[1], Mathéo BERTHET[2], Bénédicte DAYDE-CAZALS[2], Jean-François BERRET[3], Eduardo ANGLES CANO[*,1]

[1]*Faculté de Santé, Université Paris Cité, INSERM, Optimisation thérapeutique en neuropharmacologie, U1144, 75006, Paris, France.*
[2]*Specific Polymers, ZAC Via Domitia, 150 Avenue des Cocardières, 34160 Castries, France.*
[3]*Université Paris Cité, CNRS, Matière et Systèmes Complexes, 75013 Paris, France.*

[§] H.D. and R.A.R. contributed equally to this work. *Email: Eduardo.Angles-Cano@inserm.fr

### Supplemental Method-1 Coating protocol

To investigate the interaction and phase behavior of cerium oxide nanoparticles (CNPs) with phosphonic acid-based functional polymers, we used the method of continuous variation in combination with electrostatic complexation principles (1-3). Both polymer and CNP dispersions were prepared under identical acidic conditions (pH 1.4) and at matched concentrations, here $c_0$ = 1 g L-1. They were then mixed across a broad range of volumetric ratios (X), from $10^{-3}$ to $10^3$, covering polymer-rich (X≪1) to nanoparticle-rich (X≫1) conditions. A key observation from this approach was the identification of a critical mixing ratio, $X_C$, below which the resulting nanoparticle–polymer complexes remained colloidally stable across a broad pH window (1 < pH < 9) (2). In contrast, mixtures with $X > X_C$ exhibited aggregation or precipitation. This destabilization is attributed to incomplete surface coverage and weak screening of the van der Waals attractions. The final preparation steps for in vitro use of the CNPs involved filtration through a 0.22 μm cellulose acetate membrane, concentration of the dispersions using a 50 kDa cut-off AMICON filter to achieve a final concentration of 20–50 g L$^{-1}$, followed by autoclaving at 120 °C and 2 × 10$^5$ Pa for 2 hours. The resulting dispersions were stored at 4 °C.

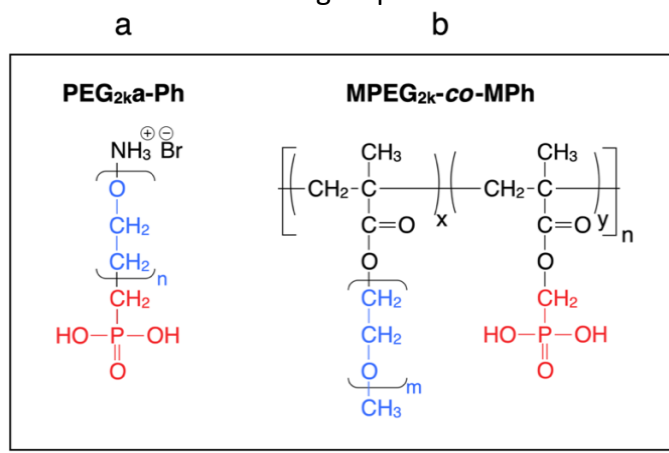

**Supplementary Figure -1**: **Chemical structures and abbreviations of the polymers used in this work** (2). **(a)** Poly(ethylene glycol) with the molecular weight 2 kDa terminated with a single phosphonic acid on one side and with a tertiary amine on the other. The polymer is abbreviated as PEG$_{2k}$a-Ph. **b)** Statistical copolymer MPEG$_{2k}$-*co*-

MPh made from MPEG$_{2k}$ and MPh monomers, wherein MPEG$_{2k}$ refers to a PEG methacrylate macromonomer with PEG molecular weight of 2 kDa and MPh to a methacrylate monomer bearing a phosphonic acid functional group. The comonomer proportions are listed in Supplementary Table I.

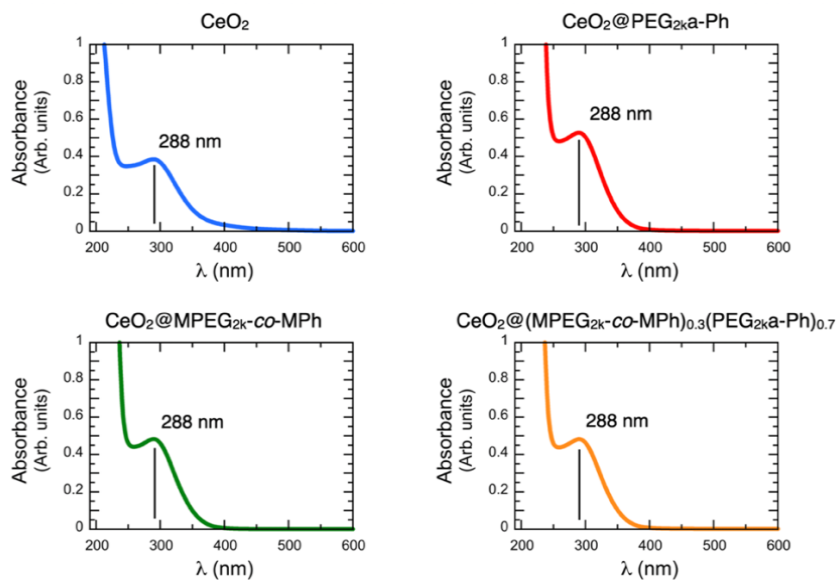

**Supplementary Figure 2. UV-visible spectra of bare and coated nanoparticles.**
*Representative absorbance spectra of bare and polymer-coated cerium oxide nanoparticles (CNPs), diluted to concentrations between $2 \times 10^{-2}$ and $4 \times 10^{-2}$ g L$^{-1}$, exhibit a characteristic peak at 288 nm. This peak, attributed to cerium oxide, remains unchanged upon coating. According to Beer–Lambert's law, the absorbance at 288 nm can be used to determine nanoparticle concentration in dispersion.*

## Supplemental Method-2 **Polymer coating**

The polymers used for coating oxide nanoparticles were synthesized by Specific Polymers via free radical polymerization. They are functional polymers featuring two key components: a phosphonic acid group for anchoring to the positively charged CeO$_2$ surface, and a polyethylene glycol chain for stability and preventing protein adsorption. For targeting NETs, two different polymers were synthesized: i) one homopolymer terminated by a single acidic phosphonic acid on one side and a primary amine at the PEG end, denoted PEG$_{2k}$a-Ph, and ii) a random copolymer referred to as MPEG$_{2k}$-co-MPh, where MPEG$_{2k}$ denotes a PEG methacrylate monomer with a PEG molecular weight of 2 kDa, and MPh stands for a methacrylate monomer bearing a phosphonic acid functional group. The polymer structures are provided in Supplementary Figure 1. Of note, for statistical copolymers, surface anchoring is strengthened by the presence of multiple phosphonic acid groups per polymer chain (Supplementary Table I). The resulting coating particles are designated CeO$_2$@PEG$_{2k}$a-Ph, CeO$_2$@MPEG$_{2k}$-co-MPh and CeO$_2$@ MPEG$_{2k}$-co-MPEG$_{2k}$ka-co-MPh, respectively.

| Polymers | $M_w^{Pol}$ (g mol$^{-1}$) | $M_n^{Pol}$ (g mol$^{-1}$) | $x - y$ | Phosphonic acids/polymer |
|---|---|---|---|---|
| PEG$_{2k}$a-Ph | 2000 | 2000 | - | 1.0 |
| MPEG$_{2k}$-*co*-MPh | 62300 | 21750 | x = 0.48 - y = 0.52 | 9.8 |

**Supplementary Table I:** *Molecular characteristics of the phosphonic acid PEGylated polymers and copolymers synthesized in this work. The chemical structures of the functional polymers along with the definitions of the comonomer proportions x and y are provided in Supplementary Figure 1.*

The nomenclature used here is identical to that of the review article on the application of these functional polymers to nanomedicine (2). For simplicity, CNPs coated with the above functional polymers will also be referred to as PEGylated CNPs. Due to the presence of an amine function at the end of the PEG chain, particles coated with PEG2ka-Ph are positively charged. In contrast, CNPs coated with the $CeO_2$@MPEG$_{2k}$-*co*-MPh are neutral. Using dynamic light scattering, it was verified that the coating layer thickness was of the order of 5-10 nm, in line with partially stretched PEG chain dimensions (1).

| Nanoparticles | $D_H$ (nm) | $h$ (nm) | $\zeta$ (mV) | Polymers per particle | PEG density (nm$^{-2}$) |
|---|---|---|---|---|---|
| Bare $CeO_2$ | 9.0 | 0 | +21 ± 1 | 0 | 0 |
| $CeO_2$@PEG$_{2k}$a-Ph | 18.1 | 4.5 | +27.7 ± 0.6 | 103 | 0.38 |
| $CeO_2$@MPEG$_{2k}$-MPh | 27.2 | 9.1 | +1.4 ± 0.1 | 15 | 0.28 |
| $CeO_2$@MPEG$_{2k}$-MPEGa$_{2k}$-MPh | 31.5 | 11.2 | +5.8 ± 0.3 | 26 | 0.62 |

**Supplementary Table II:** *Hydrodynamic diameter ($D_H$), polymer brush thickness (h), zeta potential ($\zeta$) determined for polymer coated nanoparticles. The two last columns are the number of polymers per particle and the PEG density as determined from complementary measurements (1, 4).*

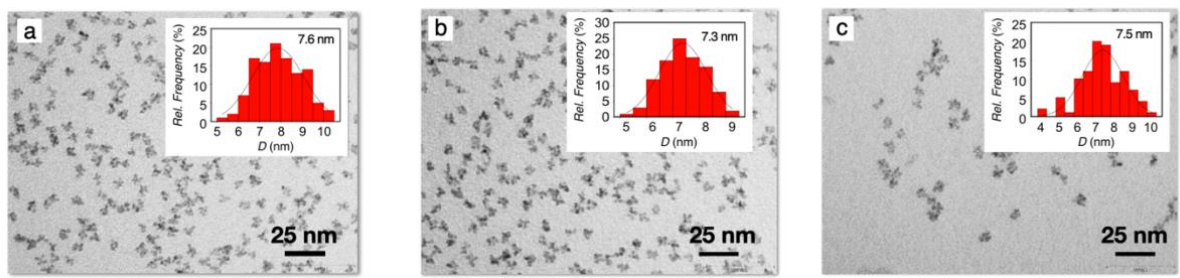

**Supplementary Table III:** *Transmission electron microscopy images of cerium oxide nanoparticles: a) $CeO_2$, b) MPEG$_{2k}$-co-MPh and c) $CeO_2$@MPEG$_{2k}$-MPEGa$_{2k}$-MPh. Insets: size distributions obtained for n = 100 particles. The similarity between the TEM sizes of coated and uncoated particles arises from the very low electron contrast of the polymer shells, which makes them invisible in TEM. However, the presence and thickness of these polymer brushes can be determined by light scattering experiments.*

## Supplementary Figure 3 Characterization of NETs in terms of DNA and elastase

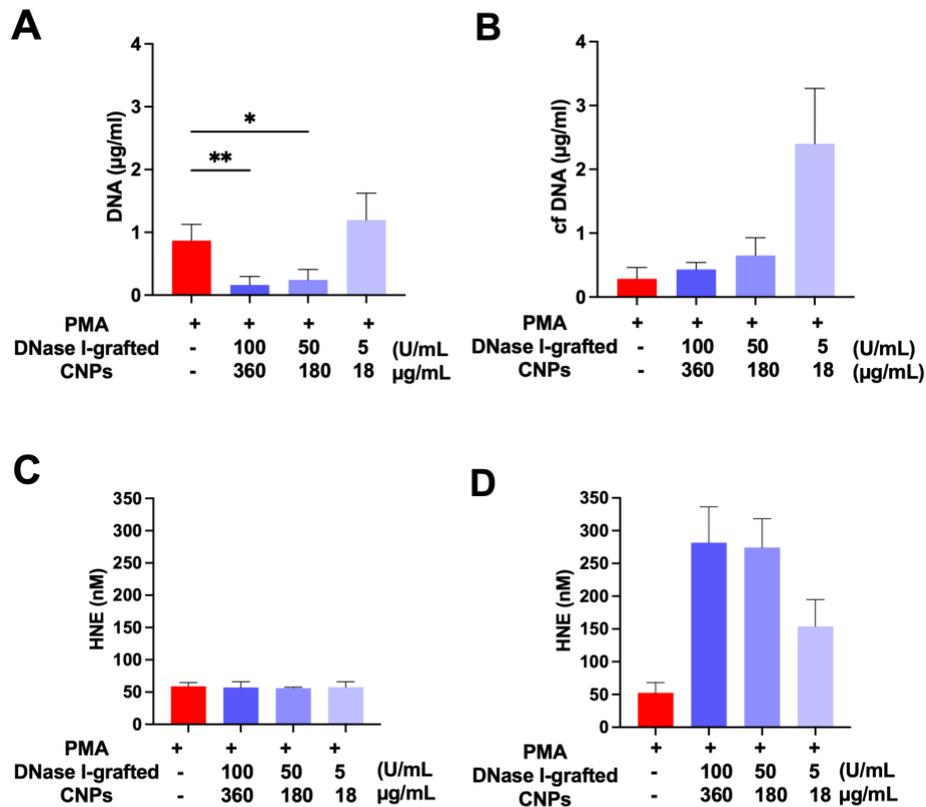

**Supplementary Figure 3 Identification of NETs by quantification of DNA and elastase content.**
*Neutrophils were isolated from human blood and seeded onto a 96-well plate at a density of 1 x 10⁵ cells/well. The neutrophils were then incubated with 50 nM PMA (red bars) or buffer alone (HBSS, green bars). The DNase I-grafted nanoparticles were added 30 minutes before PMA stimulation at different concentrations of CNP/DNase I ratio (360 μg/100 U, 180 μg/50 U, and 18 μg/5 U per mL (purple bars from darkest to lightest, respectively). NETs formation was assessed by measuring two key components, DNA using a fluorescent dye (SYBR Green) and human neutrophil elastase (HNE) using a selective chromogenic substrate. Data analysis and graphing were performed with the Prism software. Panels **A** and **B** represent DNA quantification, while panels **C** and **D** show HNE quantification. More specifically the measurements from the wells' bottom are shown in **A** and **C**, while the measurements from the corresponding supernatants are shown in **B** and **D**. cfDNA: cell-free DNA. Error bars represent SD, n=5, statistical significance is indicated by asterisk \*P<0.01, \*\*P<0.004.*

**Supplementary Figure-4 Grafting of CY5 onto CNPs (CeO$_2$@(MPEG$_{2k}$-co-MPh)$_{0.3}$(PEG$_{2k}$a-Ph)$_{0.7}$ ) Cy5 (NHS ester) was used to validate DiOC6 adsorption onto CNPs via electrostatically driven physisorption, chosen to avoid competition with DNase I immobilization.** Supplementary Figure **4 shows intracellular localization of DNA and CY5 (merged images, right panel).**

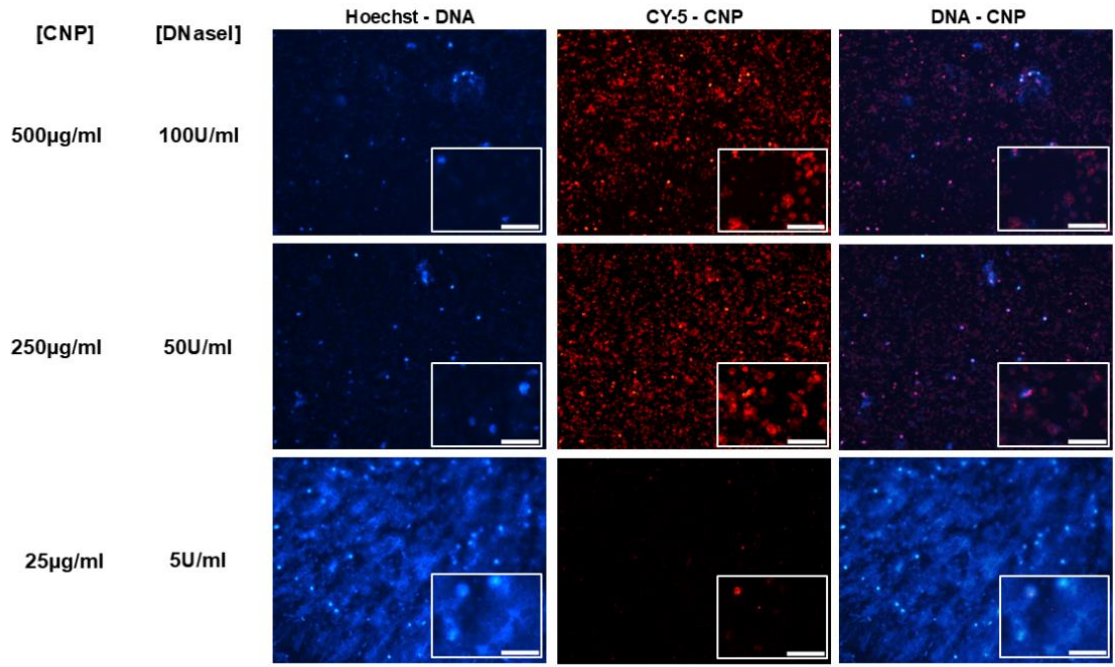

**Supplementary Figure 4 Impact and detection of CeO$_2$@(MPEG$_{2k}$-co-MPh)$_{0.3}$(PEG$_{2k}$a-Ph)$_{0.7}$ grafted with DNase-I and CY5.** DNase-I and Cyanine-5 NHS ester (CY5) were immobilized on CNPs as indicated in main text: **Grafting of DNase-I.**

*Neutrophils were isolated from human blood and seeded onto a 96-well plate at a density of 1 x 10$^5$ cells/well. The cells were then incubated with DNase I-CY5-grafted nanoparticles at various concentrations: (**A**) 100U/mL, (**B**) 50 U/mL, and (**C**) 5 U/mL for 30 minutes. This stimulation was followed by the addition of 50nM PMA to induce the formation of NETs. The CNP/DNase I concentration per mL were 500 μg/100 U, 250 μg/50 U, and 25μg/5 U. The images shown in **A**, **B** & **C** were detected in optic fields of three wells for each condition using the x10 objective of a Zeiss AxioObserver D1 fluorescence microscope equipped with a CCD Imaging camera. n=5. Representative images of fluorescence microscopy using Hoechst 33342 for DNA staining (left column), and CY5 CNP labelling (middle raw). Merged images of the left and middle columns are shown in the right column. Inset (500 % zoom, scale bar 50μM).*

**Supplementary Figure-5 CNP antioxidant activity. Coating effect on NETs formation**

We tested CNPs coated with various PEG polymers to identify formulations that retain their antioxidant activity in the presence of neutrophils

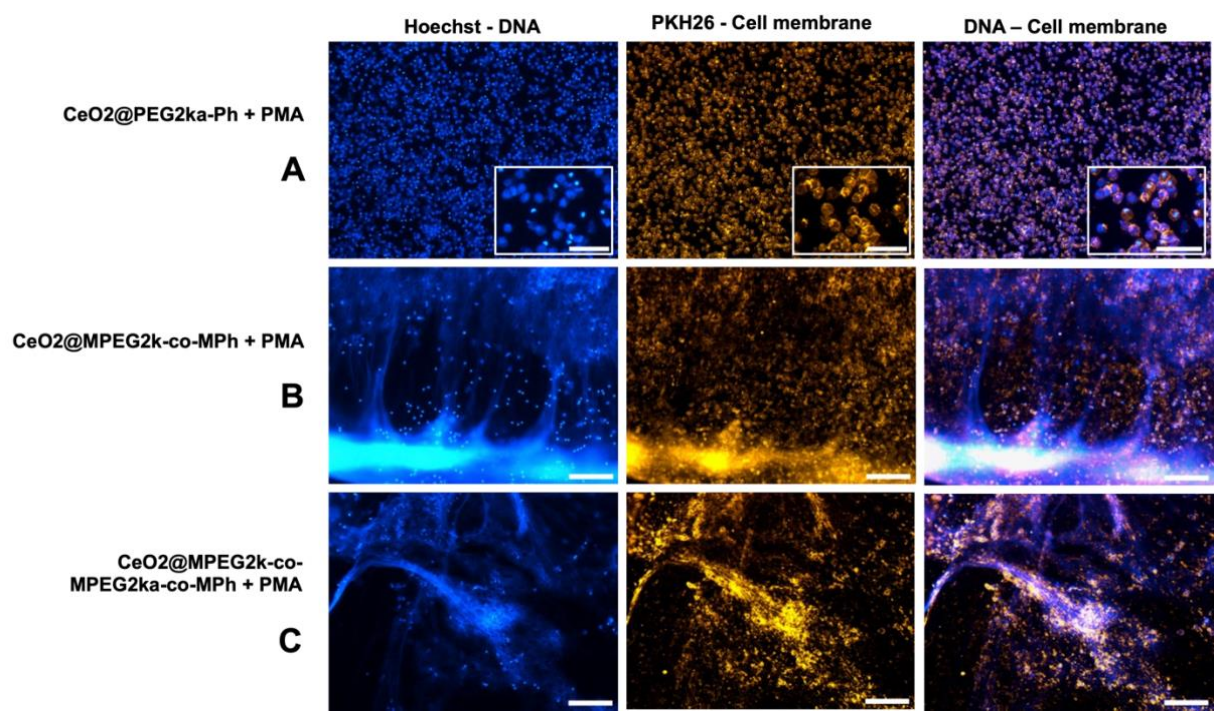

**Supplementary Figure 5. Impact of cerium oxide nanoparticles on PMA-stimulated neutrophils.**

*Neutrophils isolated from human blood were seeded on a 96-well plate at $1x10^5$ cells/well as indicated in Figure 1. Nanoparticles coated with different PEGylated polymers: (**A**) CeO$_2$@PEG$_{2k}$a-Ph, (**B**) CeO$_2$@MPEG$_{2k}$-co-MPh, (**C**) CeO$_2$@MPEG$_{2k}$-co-MPEG$_{2ka}$-co-MPh, were added at 1000µg/ml to neutrophil-containing wells 30 min prior to stimulation with 50 nM PMA (n=3). Representative images of fluorescence microscopy: Left column: Hoechst 33342 for DNA staining. Middle column: PKH26 staining for membrane labelling. Right column: merged left and middle column images. Scale bar 50 µm (**A** inset) and 150 µm (**B**, **C**). Representative neutrophil fluorescence images captured as indicated in Figure 1.*